\documentclass[twocolumn,showpacs,preprintnumbers,draftclsnofoot]{revtex4}
\usepackage{amsmath}
\usepackage{amssymb}
\usepackage{graphicx}
\usepackage{dcolumn}
\usepackage{bm}
\usepackage{amssymb}
\usepackage{graphicx}
\usepackage{dcolumn}
\usepackage{bm}

\begin{document}

\title{ Optical injection of a spin current into a zigzag nanoribbon of monolayer $MoS_{2}$ with antiferromagnetic Kekule distortion  }
\author{Ma Luo\footnote{Corresponding author:luom28@mail.sysu.edu.cn} }
\affiliation{The State Key Laboratory of Optoelectronic Materials and Technologies \\
School of Physics\\
Sun Yat-Sen University, Guangzhou, 510275, P.R. China}

\begin{abstract}

The Kekule pattern of the (anti)ferromagnetic exchange field on monolayer $MoS_{2}$ can be induced by proximity to the $(111)$ surface of $BiFeO_{3}$ on both sides. The three-band tight binding model of the $MoS_{2}$ layer with Kekule patterned exchange field is applied to describe the heterostructures. The tight binding model is justified by the first principle calculations. The magnetization orientations of the substrates control the pattern of the exchange field, which then switches the band structures of the lowest zigzag edge states between being metallic and insulating. The lowest four zigzag edge bands provide conducting channels with a spin-polarized current. Optical excitation of carriers in these bands generates sizable spin and charge currents, which are theoretically modeled by the perturbation solution of the semiconductor Bloch equation. The injected spin currents have multiple resonant peaks at a few frequencies, which can be switched off by rotating the magnetization orientations of the substrates.

\end{abstract}

\maketitle

\section{Introduction}

Transition metal dichalcogenides (TMDs) are exotic two-dimensional materials with large spin-orbit coupling (SOC) that afterwards exhibit valleytronic and spintronic physics \cite{XiaodongXu16,XiangfengDuan16}. Optical excitation of valley-polarized excitons \cite{QianNiu08} has been realized in experiments \cite{TonyFHeinz12,XiaodongCui12,JiFeng12,XiaodongXu13,WangYao14,TingYu15} and proposed for use in valleytronic devices \cite{WenYuShan15,Bertoni16}. Optical injection of a spin current into monolayer TMDs has been studied and proposed for use in spintronic devices \cite{Rodrigo15,Arzate16}. Because the band gaps of TMDs vary between 1 eV and 1.5 eV \cite{GuiBinLiu13,ShiangFang15}, the optospintronics of TMDs in the bulk is restricted to the optical frequencies within the near-infrared and visible light range. Proximity to a ferromagnetic substrate induces a spatially uniform ferromagnetic exchange field on the TMD layer, which in turn induces valley splitting \cite{NataliaCort19}. Because the inversion symmetry is broken, Rashba SOC is induced. As a result, spin mixing and canted spins are generated in the bulk and zigzag edge states. Recently, a theoretical study has suggested that the zigzag edges of TMDs carry a localized spin current with canting of the spin orientation \cite{NataliaCort19}. Similar types of localized spin current at the edge of nanostructures with Rashba SOC have been predicted \cite{KaiChang06,KaiChang11,WenKaiLou11,KaiChang14}. These systems host a spatially localized spin current as an information carrier, affording promising candidates for spintronic applications. The scheme for optical excitation of a localized spin current at the edge or topological interface of nanostructures makes integrated optospintronic devices feasible \cite{luo17}. For the zigzag nanoribbons of TMDs, optical excitation of carriers requires a small optical frequency, which could be around 0.1 eV. As a result, these nanostructures open the door for optospintronics with a far-infrared optical field as an excitation source.

The exchange field in TMD layers is critical for the property of a localized spin current. Replacing the ferromagnetic substrate by an antiferromagnetic substrate could also induce an exchange field in the TMD layer. Additionally, antiferromagnetic materials have multiple advantages over ferromagnetic materials for spintronic applications, such as the absence of parasitic stray fields and ultrafast magnetization dynamics \cite{Jungwirth16,Baltz18,Zelezny18,BGPark11}. Proximity of TMDs to the antiferromagnet $MnO$ induces a uniform exchange field \cite{LeiXu18} due to the lattice matching between the $Mo$ or $W$ atoms in the TMD layer and the $Mn$ atoms on the $(111)$ surface of $MnO$.

Kekule distortion has been studied in graphene heterostructures. Because of the outstanding physical characteristics of graphene \cite{CastroNeto09}, multiple schemes of graphene-based spintronic physics have been studied, such as optical injection of spin and valley currents \cite{Inglot14,Rioux14,Inglot15,Kaladzhyan15,Reinaldo17} and robust spin currents into topological phases \cite{CLKane051,CLKane052,TsungWeiChen11}. The Kekule distortion in graphene can be induced by proximity to a substrate with the same lattice structure but with a lattice constant $\sqrt{3}$ times that of graphene \cite{ChangYuHou07,ClaudioChamon08,DoronLBergman13,GianlucaGiovannetti15,EliasAndrade19}. One example is graphene on $In_{2}Te_{2}$ \cite{GianlucaGiovannetti15}. A $\sqrt{3}\times\sqrt{3}$ supercell of graphene is conformal with the primitive unit cell of the substrate. The band folding in the supercell brings the $K$ and $K^{\prime}$ valleys to the $\Gamma$ point in the Brillouin zone. The Kekule distortion mixes the quantum states of the two valleys, which could open a bulk gap.

In this article, the proximity effect of TMDs on the $(111)$ surface of $BiFeO_{3}$, which induces a nonuniform exchange field with the Kekule pattern, is studied. The effect changes the physical properties of the localized spin current at the zigzag edge. We consider the Kekule distortion of monolayer $MoS_{2}$, which is induced by intercalating $MoS_{2}$ between two substrates with a $(111)$ surface of $BiFeO_{3}$, i.e., the $BiFeO_{3}/MoS_{2}/BiFeO_{3}$ heterostructure. Experimental fabrication of the monolayer $MoS_{2}/BiFeO_{3}$ heterostructure is feasible \cite{YangLi18}. At room temperature, $BiFeO_{3}$ is G-type antiferromagnetic, so all $Fe$ atoms at the same $(111)$ plane have the same magnetization orientation \cite{Baettig05,Neaton06,Albrecht10,ZhenhuaQiao14}. The $Fe$ atoms on the $(111)$ surface arrange in a triangular lattice with a lattice constant of 5.50 ${\AA}$ \cite{Baettig05,Neaton06}, which is only a $0.46\%$ mismatch with $\sqrt{3}$ times the lattice constant of $MoS_{2}$, which is $a_{0}=$3.19 ${\AA}$ \cite{ZYZhu11,DiXiao12}. This article focuses on the Y-Kekule distortion. In one unit cell, the $Fe$ atom on the surface of one substrate is on top of one of the three $Mo$ atoms; the $Fe$ atom on the surface of the other substrate is on top of a different $Mo$ atom. The magnetic moments of the $Fe$ atoms induce larger and smaller exchange fields at the nearest and next nearest $Mo$ atoms, respectively. By rotating the magnetization orientation of the two substrates, the exchange fields of the three $Mo$ atoms can be changed, in turn controlling the band structure of the zigzag edge states. With the appropriate Kekule pattern of the exchange field and energy level, the forward and backward moving zigzag edge modes have opposite spins. Optical excitation of carriers to these modes generates a sizable spin current, which is localized at the $Mo$-terminated zigzag edge.

This article is organized as follows. In Section. II, the atomic structure of the $BiFeO_{3}/MoS_{2}/BiFeO_{3}$ heterostructure is described, and the effective tight-binding Hamiltonian is described. In Section. III, the band structures and spin texture of zigzag nanoribbons are discussed. In Section. IV, optical injection of a spin current is calculated and discussed. In Section. V, the conclusion is given.

\section{The Atomic Structure and Tight-Binding Model}

\begin{figure}[tbp]
\scalebox{0.66}{\includegraphics{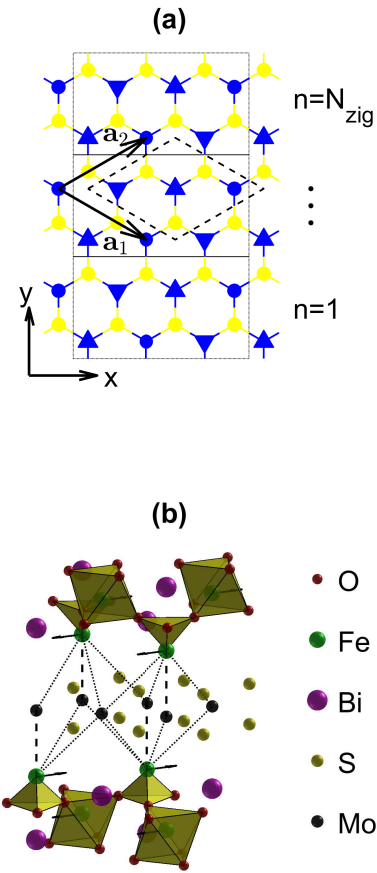}}
\caption{ (a) Lattice configuration of a $MoS_{2}$ monolayer with antiferromagnetic Kekule distortion. The yellow dots represent the $S_{2}$ sites. The blue circular dots, upward-pointing triangles, and downward-pointing triangles represent the $Mo$ sites with different magnetic exchange fields. The dashed line indicates the primitive unit cell. The dotted lines indicate the rectangular unit cells. (b) Three-dimensional atomic structure of the $BiFeO_{3}/MoS_{2}/BiFeO_{3}$ heterostructure within one rectangular unit cell. The dashed and dotted lines connect the nearest and next nearest $Fe$ and $Mo$ atoms, respectively. The arrows at the $Fe$ atoms represent the local magnetic moment.   \label{fig1}}
\end{figure}

The lattice structure of the $BiFeO_{3}/MoS_{2}/BiFeO_{3}$ heterostructure with the Y-Kekule configuration is plotted in Fig. \ref{fig1}(a), which only indicates the lattice sites of $Mo$ and $S$ atoms. The primitive unit vectors of the bulk are $\mathbf{a}_{1(2)}=\frac{3\sqrt{3}}{2}a_{0}\hat{x}\pm\frac{3}{2}a_{0}\hat{y}$, which define the primitive unit cell of the bulk. The primitive unit cell includes three $Mo$ atomic lattice sites with different magnetic exchange fields, which are indicated as different shapes of the lattice points in Fig. \ref{fig1}(a). We consider zigzag nanoribbons with translational invariance along the $\hat{x}$ direction. The rectangular unit cell is defined by the vectors $\mathbf{a}_{1}+\mathbf{a}_{2}$ and $-\mathbf{a}_{1}+\mathbf{a}_{2}$ and contains two primitive unit cells. The unit cell of the zigzag nanoribbon consists of $N_{zig}$ rectangular unit cells along the width direction, i.e., the $\hat{y}$ direction. The atomic structure within one rectangular unit cell of the $BiFeO_{3}/MoS_{2}/BiFeO_{3}$ heterostructure is plotted in Fig. \ref{fig1}(b). For better visualization, only two Fe layers in each $BiFeO_{3}$ substrate are plotted. Each $Fe$ atom on the surface is on top of a $Mo$ atom, which are connected by the dashed line in Fig. \ref{fig1}(b). The magnetic moment at the $Fe$ atoms induces a magnetic exchange field at the $Mo$ atoms. The strength of the exchange field depends on the distance between the $Fe$ and $Mo$ atoms. Thus, the exchange field at the $Mo$ atoms on top of an $Fe$ atom is larger than the exchange field at the other $Mo$ atoms that are not on top of an $Fe$ atom. We assume that the $Fe$ atoms of the bottom (top) $BiFeO_{3}$ substrate are on top of the $Mo$ atomic lattice sites plotted as upward-pointing (downward-pointing) triangles in Fig. \ref{fig1}(a). Thus, the bottom (top) $BiFeO_{3}$ substrate induces exchange field $\mathbf{M}_{B}$ ($\mathbf{M}_{T}$) at the upward-pointing (downward-pointing) triangle lattice sites and $\lambda\mathbf{M}_{B}$ ($\lambda\mathbf{M}_{T}$) at the other lattice sites. The parameter $\lambda$ characterizes the degree of nonuniformity of the exchange field induced by each substrate. If the substrate induces a nearly uniform exchange field, $\lambda$ is nearly one; in contrast, if the substrate induces a highly nonuniform exchange field, $\lambda$ is nearly zero. Combining the exchange fields from both substrates, the exchange field at the lattice sites plotted as upward-pointing triangles, downward-pointing triangles, and circular dots are $\mathbf{M}_{\bigtriangleup}=\mathbf{M}_{B}+\lambda\mathbf{M}_{T}$, $\mathbf{M}_{\bigtriangledown}=\lambda\mathbf{M}_{B}+\mathbf{M}_{T}$, and $\mathbf{M}_{\bigcirc}=\lambda\mathbf{M}_{B}+\lambda\mathbf{M}_{T}$, respectively. The strengths of the exchange fields from the top and bottom substrates are the same, which are designated as $B_{0}=|\mathbf{M}_{B}|=|\mathbf{M}_{T}|$.

The heterostructure is modeled by the tight-binding Hamiltonian $H=H_{MoS_{2}}+H_{ex}+H_{R}$. The Hamiltonian of monolayer $MoS_{2}$ is given by the three-band tight-binding model with the three d orbital basis $\{|d_{z^{2}},s\rangle,|d_{xy},s\rangle,|d_{x^{2}-y^{2}},s\rangle\}$ and spin index $s=\pm1$. The Hamiltonian of the Rashba SOC $H_{R}$ is given by the intrasite inter-$\{|d_{xy},s\rangle,|d_{x^{2}-y^{2}},s\rangle\}$-orbital mixing matrix for each lattice site. The Rashba coupling strength is assumed to be $\lambda_{R}=76$ meV. The detailed form of the Hamiltonians $H_{MoS_{2}}$ and $H_{R}$ can be found in Ref. \cite{GuiBinLiu13} and \cite{NataliaCort19}, respectively. The Hamiltonian of the exchange field is given as $H_{ex}=\sum_{j}M_{cvv}\otimes(\mathbf{M}_{j}\cdot\mathbf{\sigma})$, with the summation covering all lattice sites,  $M_{cvv}=diag\{1,0.8252,0.8252\}$, $\mathbf{\sigma}=\hat{x}\sigma_{x}+\hat{y}\sigma_{y}+\hat{z}\sigma_{z}$ and $\sigma_{x,y,z}$ being the Pauli matrix of spin. The diagonal matrix elements of $M_{cvv}$ are the magnetic coupling coefficient to the conduction and valence bands.

\begin{figure}[tbp]
\scalebox{0.439}{\includegraphics{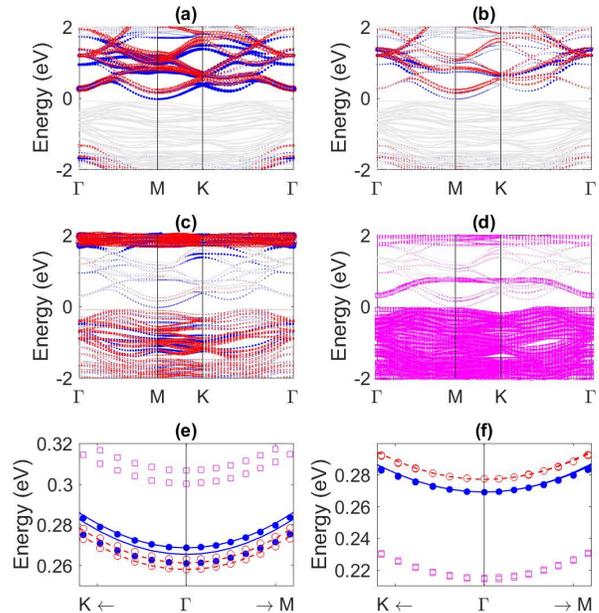}}
\caption{ Orbital projected band structures for the $BiFeO_{3}/MoS_{2}/BiFeO_{3}$ heterostructure with the Y-Kekule configuration obtained from DFT calculations are plotted in (a-d). The Fermi energy is set to zero. The symbol size is proportional to its projection in the corresponding state. (a) Contribution from the three $Mo$ $d$ orbitals $(d_{z^{2}},d_{xy},d_{x^{2}-y^{2}})$. (b) Contribution from the other orbitals of $MoS_{2}$. (c) Contribution from $Fe$ atoms. In (a-c), spin up and down states are plotted as solid (blue) and empty (red) dots, respectively. (d) Contribution from ($Bi$, $O$, $H$) atoms, with both spins being plotted as empty (magenta) squares. The band structures are zoomed in on around the $\Gamma$ point for the Y-Kekule and O-Kekule configurations in (e) and (f), respectively. Only contributions from the three $Mo$ $d$ orbitals and ($Bi$, $O$, $H$) atoms are included in (e-f). The solid (blue) and dashed (red) lines in (e-f) are fittings of the spin up and down band structures by the tight-binding model, respectively. \label{figBVASP}}
\end{figure}

To justify the tight-binding model with the Kekule pattern of the exchange field, we performed density functional theory (DFT) calculations based on the projector augmented wave method as implemented in the Vienna ab initio simulation package (VASP) \cite{vasp001,vasp002}. The generalized gradient approximation exchange-correlation functional \cite{vasp003} was used. The SOC was neglected. The spin polarizations were collinear. The Hubbard U was chosen to be 4 eV for the $d$ orbital of the Fe atoms. The lattice constant of the $BiFeO_{3}/MoS_{2}/BiFeO_{3}$ heterostructure was chosen to be 5.50 ${\AA}$, which corresponds to the lattice constant of $a_{0}=3.175$ ${\AA}$ for the $MoS_{2}$ monolayer. The kinetic energy cutoff was set to 550 eV. Each $BiFeO_{3}$ substrate contains six layers of $Fe$ atoms along the $(111)$ direction. The $Fe$ atoms in adjacent layers have opposite spin polarizations. The $Fe$ atoms in proximity to the top and bottom surfaces of the $MoS_{2}$ monolayer have the same spin polarizations, corresponding to the tight-binding model with $\mathbf{M}_{T}=\mathbf{M}_{B}$. The $Bi$ atoms at the open boundaries are passivated by hydrogen atoms. The heterostructures with both the Y-Kekule and O-Kekule configurations were calculated. The initial atomic structure of the O-Kekule configuration was obtained from that of the Y-Kekule configuration by shifting the atoms in the $MoS_{2}$ monolayer along the $+\hat{y}$ direction by 1.833 ${\AA}$ while fixing the atoms in the $BiFeO_{3}$ substrates. After structural relaxations in which the forces converged to less than 0.01 eV/${\AA}$, the interlayer distances between the $MoS_{2}$ monolayer and the $Fe$ $(111)$ surface were found to be 3.89 ${\AA}$ and 4.1 ${\AA}$ for the Y-Kekule and O-Kekule configurations, respectively.


The band structures of the heterostructures in the bulk with both types of Kekule configuration were calculated. The band structures of the Y-Kekule configuration are plotted in Fig. \ref{figBVASP}(a-d) within the energy range of $[-2,2]$ eV. At this energy scale, the band structures of the two Kekule configurations are nearly the same. Because the heterostructure contains a $\sqrt{3}\times\sqrt{3}$ supercell of the $MoS_{2}$ monolayer, the conduction band minima of pristine $MoS_{2}$ monolayer at the $K$ and $K^{\prime}$ points are folded into the $\Gamma$ point. Although the band gap of the heterostructure is 0.09 eV, the band gap of the localized band structure that is the projection to the $MoS_{2}$ monolayer is 1.67 eV, as shown in Fig. \ref{figBVASP}(a) and (b). Most of the quantum states in the $MoS_{2}$ monolayer are projected into the three $d$ orbitals $(d_{z^{2}},d_{xy},d_{x^{2}-y^{2}})$, as shown in Fig. \ref{figBVASP}(a). The projections to the $Fe$ ($Bi$, $O$, $H$) atoms are mainly distributed in the conduction band (valence band) above 1.8 eV (below 0 eV), as shown in Fig. \ref{figBVASP}(c) and (d), which have small mixing with the conduction band minima of the $MoS_{2}$ monolayer. The projections to the ($Bi$, $O$) atoms have large weights in two bands above the Fermi energy, whose energy range overlaps with the conduction band minima of the $MoS_{2}$ monolayer, as shown in Fig. \ref{figBVASP}(d). By zooming in on the band structure near to the $\Gamma$ point, as shown in Fig. \ref{figBVASP}(e), the projections to the $MoS_{2}$ monolayer and the ($Bi$, $O$) atoms are completely separated into two groups of bands. Because the groups of ($Bi$, $O$) atoms and the $MoS_{2}$ monolayer are spatially separated by the $Fe$ atoms on the surface of the substrates, the quantum states in the $MoS_{2}$ monolayer and the ($Bi$, $O$) atoms are weakly mixed. As a result, the conduction band minima of the $MoS_{2}$ monolayer can be well described by the three-orbital tight-binding model, which neglects the mixing with the $BiFeO_{3}$ substrate. A recent experimental study of the $MoS_{2}$ bilayer showed that the quantum states in the conduction band minima are localized within each layer with the absence of interlayer tunneling \cite{RiccardoPisoni19}. Thus, neglecting the orbital mixing between $MoS_{2}$ and $BiFeO_{3}$ in the $BiFeO_{3}/MoS_{2}/BiFeO_{3}$ heterostructure is reasonable.

The conduction band minima is fitted by the tight-binding model with a nonuniform exchange field, i.e., $B_{0}=4.3$ meV and $\lambda=0.25$, and without SOC, as shown in Fig. \ref{figBVASP}(e). The fitting is qualitatively correct because the two conduction band minima for each spin are not degenerate. By contrast, for the heterostructure in the O-Kekule configuration, the conduction band minima are fitted by the tight-binding model with a uniform exchange field, i.e., $B_{0}=2$ meV and $\lambda=1$, as shown in Fig. \ref{figBVASP}(f), as the two conduction band minima for each spin are degenerate. To demonstrate the qualitative features of the model, we assume stronger exchange coupling by using the parameters of $B_{0}=125$ meV and $\lambda=0.25$ in the remainder of this article.

\section{The Band Structure of Zigzag Nanoribbons}

\begin{figure}[tbp]
\scalebox{0.4}{\includegraphics{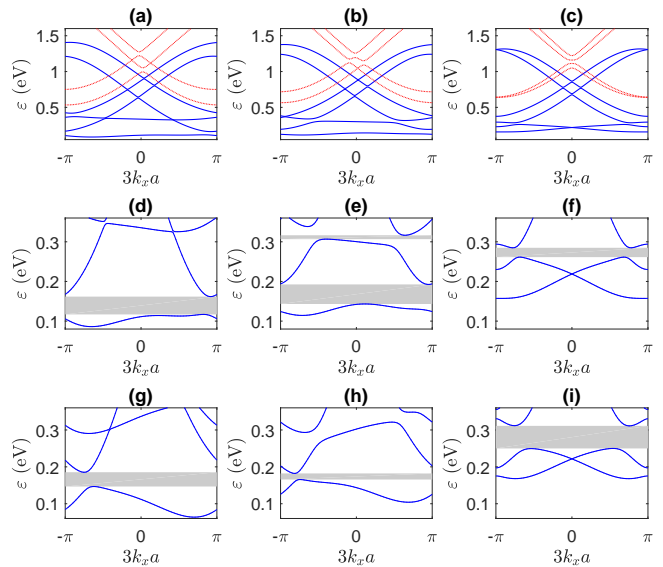}}
\caption{ The band structure of a zigzag nanoribbon with a width of $5\sqrt{3}a_{0}$. The edge states localized on the $Mo$- and $S$-edge terminations are plotted as solid (blue) and dashed (red) lines, respectively. The orientations of the magnetizations in the substrates are (a,d) $\mathbf{M}_{B}=\mathbf{M}_{T}=B_{0}\hat{z}$, (b,e) $\mathbf{M}_{B}=B_{0}\hat{z}$ and $\mathbf{M}_{T}=B_{0}\hat{x}$, (c,f) $\mathbf{M}_{B}=-\mathbf{M}_{T}=B_{0}\hat{z}$, (g) $\mathbf{M}_{B}=\mathbf{M}_{T}=B_{0}\hat{y}$, (h) $\mathbf{M}_{B}=B_{0}\hat{y}$ and $\mathbf{M}_{T}=B_{0}\hat{x}$, and (i) $\mathbf{M}_{B}=-\mathbf{M}_{T}=B_{0}\hat{y}$. In figures (d-i), the possible band gaps above the first or second edge bands are indicated by the gray area.  \label{fig2}}
\end{figure}

\begin{figure*}[tbp]
\scalebox{0.61}{\includegraphics{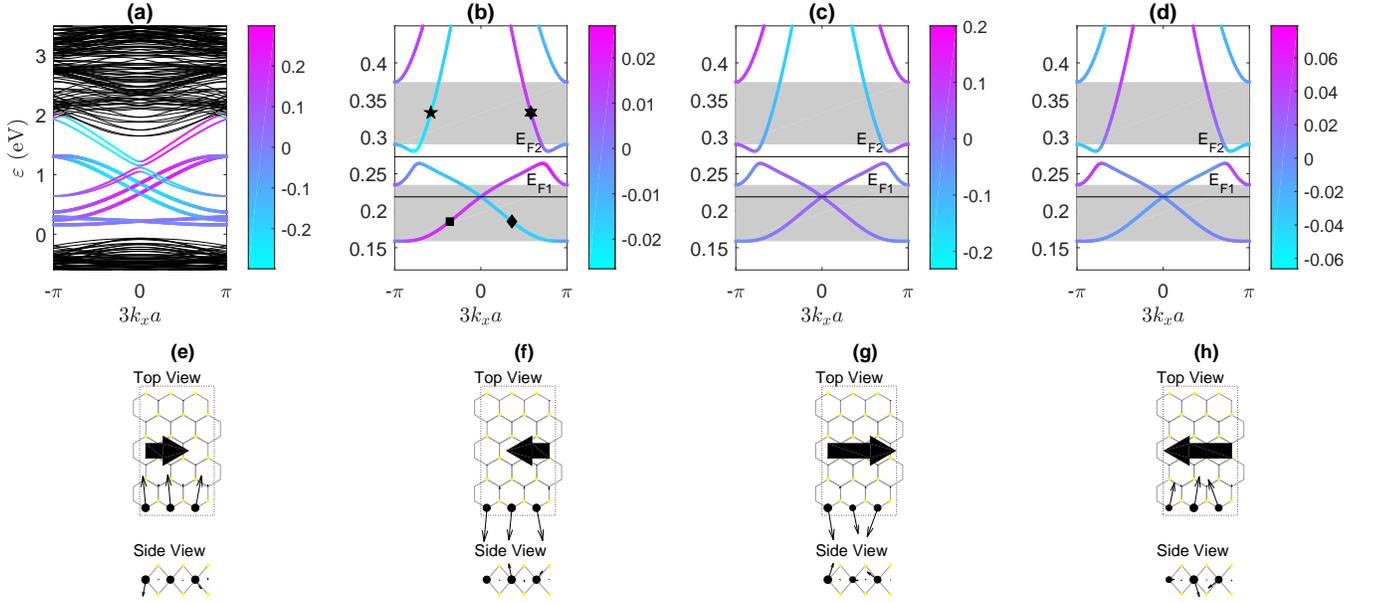}}
\caption{ The band structure of the zigzag nanoribbon with a width of $5\sqrt{3}a_{0}$ and a magnetization orientation of the substrates of $\mathbf{M}_{B}=-\mathbf{M}_{T}=B_{0}\hat{z}$ (the same as that in Fig. \ref{fig2}(c)). In (a), the bulk states are plotted as thin black lines; the edge states of the $Mo$- and $S$-edge terminations are plotted as thin and thick lines with color. The color scale corresponds to the expectation value of the velocity operator in the unit of $nm/fs$, i.e., $\langle\hat{v}_{x}\rangle$. In figures (b), (c) and (d), the energy range is zoomed in on; the color scale corresponds to the expectation value of the spin-velocity operator in the unit of $nm/fs$, i.e, $\langle\hat{s}_{x,x}\rangle$, $\langle\hat{s}_{x,y}\rangle$ and $\langle\hat{s}_{x,z}\rangle$, respectively. The physical features for the quantum states marked by a square, diamond, pentagram or hexagram dot in (b) are plotted in (e), (f), (g) and (h), respectively. Only the three unit cells near the $Mo$-edge termination are plotted. The probability density at each $Mo$ site is indicated by the size of the marker. The local spin moment at each $Mo$ site, i.e., $\langle\sigma_{x}\rangle\hat{x}+\langle\sigma_{y}\rangle\hat{y}+\langle\sigma_{z}\rangle\hat{z}$, is indicated by the arrows. The velocity $\langle\hat{v}_{x}\rangle$ is indicated by the horizontal thick arrow in the top view. \label{fig3}}
\end{figure*}

We consider zigzag nanoribbons with the width of five rectangular unit cells, i.e., $N_{zig}=5$. The band structure of the zigzag nanoribbon can be obtained by diagonalization of the tight-binding Hamiltonian with a Bloch periodic boundary along the zigzag direction. In addition to the bulk states, edge states with energy within the bulk gap are found. For the pristine $MoS_{2}$ zigzag nanoribbon, the unit cell is three times smaller than that in Fig. \ref{fig1}(a) along the zigzag direction; the band structures of the edge states are connected to the conduction bands and have a finite gap with the valence bands \cite{GuiBinLiu13}. By choosing the unit cell in Fig. \ref{fig1}(a), the two band valleys at the $K$ and $K^{\prime}$ points are folded into the $\Gamma$ point, i.e., $k_{x}=0$. In the presence of the Kekule pattern of the magnetic exchange field, the two band valleys are coupled, which modifies the band structure of the bulk and edge states. The band structures of the edge states are shown in Fig. \ref{fig2}. Because the edge states are highly localized within two unit cells beyond each zigzag edge, the edge states of the two terminations hardly overlap. Thus, the band structure of a zigzag edge with a larger width is nearly the same. If the magnetization orientation of the bottom substrate is fixed, then rotation of the magnetization orientation of the top substrate between parallel and antiparallel relative to that of the bottom substrate changes the band structures. For the edge states localized on the $Mo$-edge termination, the first edge band is gapped from the second edge band for parallel magnetization [$\mathbf{M}_{B}=\mathbf{M}_{T}$, as shown in Fig. \ref{fig2} (d) and (g)], and is connected to the second edge band for antiparallel magnetization [$\mathbf{M}_{B}=-\mathbf{M}_{T}$, as shown in Fig. \ref{fig2} (f) and (i)]. Similar features are found between the second and third edge bands. For the neutral system, the Fermi level fills up to the lowest six edge bands. With hole doping of $5/(36N_{zig})$ ($4/(36N_{zig})$) in $MoS_{2}$, the Fermi level is between the first and second (second and third) edge bands. Thus, the nanoribbon is switched between insulating and metallic by the rotation of the magnetization orientation, which can function as a spin valve \cite{CCardoso18,maluo19}. For the edge states localized on the $S$-edge termination, the gap between the lower and higher two bands is also switched on and off by the rotation of the magnetization orientation. However, if the Fermi level is within this gap, then the $Mo$-edge termination edge state are always conducting, so the system cannot function as a spin valve. The spin valve function is due to the Kekule pattern of the exchange field. If the exchange field is uniform (i.e., $\lambda=1$), the gaps in Fig. \ref{fig2} do not appear for any combination of $\mathbf{M}_{B}$ and $\mathbf{M}_{T}$.

Because of the Rashba SOC and the Kekule pattern of the exchange field, the edge state can support a one-way spin current. We focus on the system with a magnetization orientation of the substrates of $\mathbf{M}_{B}=-\mathbf{M}_{T}=B_{0}\hat{z}$ (the system in Fig. \ref{fig2}(c)). The charge and spin currents are characterized by the velocity and spin-velocity operators, which are defined as $\hat{v}_{x}=(1/\hbar)\partial H/\partial k_{x}$ and $\hat{s}_{x,\kappa}=\frac{1}{2}(\hat{v}_{x}\sigma_{\kappa}+\sigma_{\kappa}\hat{v}_{x})$, respectively, where $k_{x}$ is the wave vector along the zigzag nanoribbon and $\kappa=x,y,z$. The band structures exhibiting the expectation values of the velocity and spin-velocity operators in color scales are plotted in Fig. \ref{fig3}. The band structure is symmetric under $k_{x}\Leftrightarrow-k_{x}$. The velocity of the edge states is proportional to the slope of the band structures. Within the energy ranges marked by the gray area in Fig. \ref{fig3}(b-d), only two edge states with opposite velocities occur at each energy level. The features of four typical states within these energy ranges are plotted in Fig. \ref{fig3}(e-h). The probability densities of the states are highly localized at the $Mo$-edge termination. For the two edge states in Fig. \ref{fig3}(e) and (f), the expectation values of spin-x (spin-y, spin-z) are the same as (opposite to) each other, so the expectation values of the spin-velocity operator are opposite to (the same as) each other. For the spin-y and spin-z components, this property can be designated as a one-way spin-velocity texture. For the other two edge states in Fig. \ref{fig3}(g) and (h), a similar property with larger expectation values of the velocity operator appears, which in turn generates a larger spin current. In the absence of the Rashba SOC or the Kekule pattern of the exchange field, these energy ranges vanish. As the strength of the Rashba SOC decreases, the width of these energy ranges decreases. Optical excitation of carriers within these energy ranges could inject a large spin current with y and z components because the forward and backward traveling electrons carry spin currents with the same sign. In comparison, the spin current with the x component would be smaller because the forward and backward traveling electrons carry spin currents with opposite signs. Because the inversion symmetry is absent, the nondiagonal matrix elements of the velocity operator are not symmetric under $k_{x}\Leftrightarrow-k_{x}$. Thus, optical excitation generates different carrier populations at the forward and backward traveling edge states, then injecting a charge current.

\section{Optical Spin Injection}

The optical excitation is modeled by the semiconductor Bloch equation \cite{luo17,maluo16}. In the presence of an optical field, the Hamiltonian has an additional interaction term, given as
\begin{equation}
H_{I}=-\frac{e_{0}}{m_{0}c}\mathbf{A}(t)\cdot\mathbf{P}
\end{equation}
where $e_{0}$ is the electron charge, $m_{0}$ is the electron mass, $c$ is the speed of light and $\mathbf{A}$ is the vector potential. Under the Coulomb gauge, the electric field is given by $\mathbf{E}=-\frac{1}{c}\frac{\partial\mathbf{A}(t)}{\partial t}$. We consider a continuous-wave harmonic optical field with linear polarization along the $\hat{x}$ direction and a frequency of $\omega$, so $\mathbf{E}=Re[E_{0}e^{-i\omega t}]\hat{x}$. In a realistic experimental condition, the plane wave of the optical field is approximated by the center part of the Gaussian beam with a beam width of one wavelength in vacuum, so the relation between the power of the Gaussian beam and the amplitude of the electric field is $P_{0}=\frac{|E_{0}|^{2}\pi}{4Z_{0}}(\frac{1240}{\hbar\omega})^{2}$, with $Z_{0}=376.73\Omega$ being the impedance of free space. We assume $P_{0}=10^{-5}$ W, so $E_{0}$ is a function of the optical frequency in our numerical simulation. The momentum operator along the $\hat{x}$ direction is given by the velocity operator as $P_{x}=m_{0}\partial H/\partial P_{x}=\frac{m_{0}}{\hbar}\partial H/\partial k_{x}$. Thus, the interaction Hamiltonian is given as
\begin{equation}
H_{I}=i\frac{e_{0}E_{0}}{\hbar\omega}\frac{\partial H}{\partial k_{x}}e^{-i\omega t}-i\frac{e_{0}E_{0}}{\hbar\omega}\frac{\partial H}{\partial k_{x}}e^{i\omega t}
\end{equation}
The time evolution of the density matrix obeys the semiconductor Bloch equation with the relaxation time approximation, given as
\begin{equation}
i\hbar\frac{\partial\rho(t)}{\partial t}=[\rho(t),H+H_{I}]-\frac{\hbar}{\tau}[\rho(t)-\rho^{(0)}(t)]
\end{equation}
where $\tau$ is the relaxation time of the edge states. We assume $\tau=1$ ps. The perturbation solution can be expanded as $\rho(t)=\rho^{(0)}+\rho^{(1)}(t)+\rho^{(2)}(t)+\cdots$, with $\rho^{(0)}$ being a diagonal matrix whose elements are given by the Fermi-Dirac distribution at temperature $T$. The second-order perturbation solution $\rho^{(2)}(t)$ includes second harmonic terms with time-dependent factor $e^{\pm2i\omega t}$ and zero harmonic terms $\rho^{(2)}_{0}$ that are independent of time. The injection of direct charge and spin currents is determined by the expectation of the velocity and spin-velocity operators mutiplied by the zero harmonic terms $\rho^{(2)}(t)$, given as
\begin{equation}
I_{c}=\frac{e_{0}}{3N_{k}a_{0}}\sum_{k,n,n^{\prime}}\langle k,n|\hat{v}_{x}|k,n^{\prime}\rangle\rho^{(2)}_{0,k,n^{\prime},n}
\end{equation}
and
\begin{equation}
I_{s}^{\kappa}=\frac{e_{0}}{3N_{k}a_{0}}\sum_{k,n,n^{\prime}}\langle k,n|\hat{s}_{x,\kappa}|k,n^{\prime}\rangle\rho^{(2)}_{0,k,n^{\prime},n}
\end{equation}
where $N_{k}$ is the number of sampling point of the Bloch wave number $k$. The definition of the spin currents in Eq. (5) is normalized by a factor $(\hbar/e_{0})$, so that they have the same unit as charge current.

\begin{figure}[tbp]
\scalebox{0.56}{\includegraphics{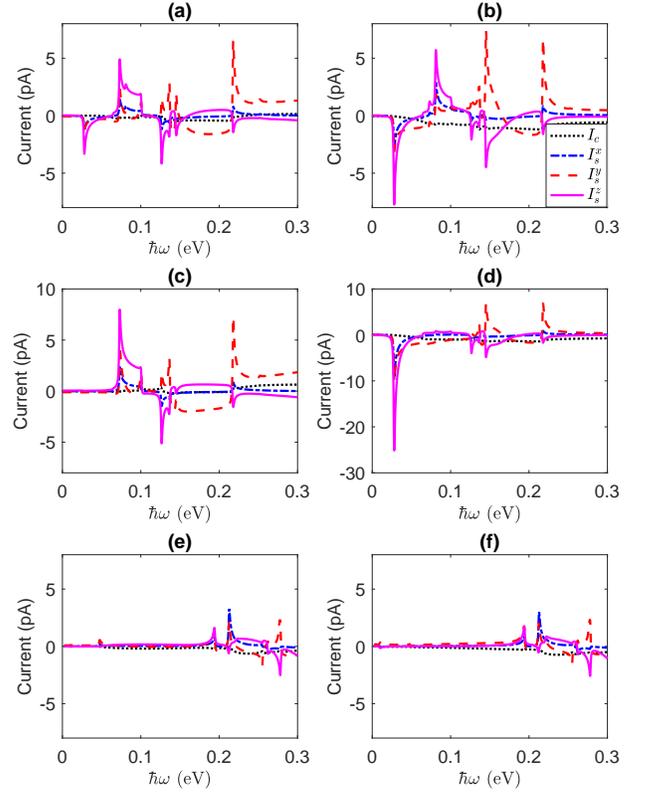}}
\caption{ Optical injection of charge and spin currents into the nanoribbon with a magnetization orientation of the substrates of $\mathbf{M}_{B}=-\mathbf{M}_{T}=B_{0}\hat{z}$ in (a-d) or $\mathbf{M}_{B}=\mathbf{M}_{T}=B_{0}\hat{z}$ in (e,f). The Fermi energy level is $E_{F1}$ in (a,c,e) and $E_{F2}$ in (b,d,f). The temperature in (a,b) and (e,f) is 300 K, and that in (c,d) is 70 K. \label{fig4}}
\end{figure}

The optical injection of charge and spin currents into the system with a magnetization orientation of the substrates of $\mathbf{M}_{B}=-\mathbf{M}_{T}=B_{0}\hat{z}$ (the same as the system in Fig. \ref{fig2}(c) and Fig. \ref{fig3}) versus the frequency of the optical field is plotted in Fig. \ref{fig4}(a) and (b), with a Fermi energy level of $E_{F1}=0.2188$ eV and $E_{F2}=0.2731$ eV, respectively, and a temperature of 300 K. The charge current is small but nonzero because the excited populations of edge states at $k_{x}$ and $-k_{x}$ at the same energy level are different. In most of the optical frequency range, $I_{s}^{y,z}$ is larger than $I_{s}^{x}$, which confirms the inference in the previous section. The spin currents peak at several resonant frequencies, which is due to the competition between the spin currents generated by the edge states within or outside of the energy ranges with a one-way spin-velocity texture (those in the gray area in Fig. \ref{fig3}(b-d)). For example, for the first resonant peak at optical frequency $\hbar\omega=0.028$ eV, the optically excited electrons (holes) are populated around the band crossing above (below) $E_{F2}$. As the optical frequency decreases, no electron is populated within the energy ranges in the gray area, so the magnitude of the spin currents sharply decreases. As the optical frequency increases, more holes are populated outside the energy ranges in the gray area, which cancel the total spin currents; thus, the magnitude of the spin current smoothly decreases. In Fig. \ref{fig4}(a), although the energy levels around the band crossing at $E_{F2}$ are above the Fermi level $E_{F1}$, the equilibrium populations (i.e., diagonal terms of $\rho^{(0)}$) in these energy levels are sizable because of the thermal excitation at room temperature. Thus, the optical field can excite electrons and holes at the upper and lower parts of the band crossing, respectively. Therefore, a resonant peak exists. If the Fermi level is raised to $E_{F2}$, as in Fig. \ref{fig4}(b), then the equilibrium population contrast between the upper and lower parts of the band crossing is larger, so the amplitude of the resonant peak becomes larger. For the same systems in Fig. \ref{fig4}(a) and (b), if the temperature is decreased to 70 K, then the injected currents change to those shown in Fig. \ref{fig4}(c) and (d), respectively. For the system in Fig. \ref{fig4}(c), the equilibrium population at the band crossing vanishes, as the energy levels around the band crossing are more than $k_{B}T$ above the Fermi level $E_{F1}$; thus, the first resonant peak disappears. For the system in Fig. \ref{fig4}(d), the equilibrium population contrast between the energy levels above and below the Fermi level $E_{F2}$ is further increased, so the magnitude of the first resonant peak is larger than that in Fig. \ref{fig4}(b). In comparison, for a suspended $MoS_{2}$ nanoribbon without an exchange field, the spin currents have no resonant peak versus the optical frequency, and the maximum spin current is less than 0.1 pA.

If the magnetization orientation of the substrates is switched to be parallel, i.e., $\mathbf{M}_{B}=\mathbf{M}_{T}=B_{0}\hat{z}$, then the optical injection of charge and spin currents is as plotted in Fig. \ref{fig4}(e) and (f). In these cases, the first few resonant peaks are switched off. The spin currents at the corresponding optical frequency are less than 0.1 pA. A few resonant peaks at a larger optical frequency with a smaller magnitude appear. This property can function as an optical spin valve because the spin currents are switched on and off by the rotation of the magnetization orientation. As the parameters of the heterostructure ($B_{0}$, $\lambda$ and $\lambda_{R}$) change, the energy ranges with a one-way spin-velocity texture change, in turn changing the resonant frequencies and magnitudes of the peaks of the spin current. As the strength of the Rashba SOC decreases, the energy ranges become smaller, so the magnitude of the resonant peaks decreases.

\section{Conclusion}

In conclusion, intercalation of monolayer $MoS_{2}$ between two $BiFeO_{3}$ substrates can induce the Kekule pattern of the exchange field in the $MoS_{2}$ layer, which is controlled by the magnetization orientation of the substrates. By employing the tight-binding model, the numerical simulations reveal that the band structures and spin texture of the zigzag nanoribbon are dependent on the nonuniform pattern of the exchange field. The edge bands can be switched between metallic and insulating. With an antiparallel magnetization orientation of the substrates, large energy ranges with a one-way spin-velocity texture for spin-y and spin-z components are found. This property enhances the optical injection of a spin current. When the optical excitation generates the maximum population of edge states with a one-way spin-velocity, the injected spin current peaks. If the magnetization orientation of the substrates is switched to be parallel, then the peaks are turned off because the band structures and spin texture are changed. As a result, the optical excitation of a localized spin current at the zigzag edge is controlled by the magnetization orientation of the substrates. Because the spin current is highly localized at the $Mo$-edge termination, experimental implementation of the scheme could be performed along a wide $MoS_{2}$ monolayer with a smooth zigzag $Mo$ edge, rather than a narrow $MoS_{2}$ zigzag nanoribbon \cite{XiaofeiLiu13}.

\begin{acknowledgments}
This project is supported by the National Natural Science Foundation of China (Grant:
11704419).
\end{acknowledgments}

\clearpage


\section*{References}

\begin{thebibliography}{99}


\bibitem{XiaodongXu16} John R. Schaibley, Hongyi Yu, Genevieve Clark, Pasqual Rivera, Jason S. Ross, Kyle L. Seyler, Wang Yao and Xiaodong Xu, Nature Reviews Materials, 1, 16055(2016).

\bibitem{XiangfengDuan16} Yuan Liu, Nathan O. Weiss, Xidong Duan, Hung-Chieh Cheng, Yu Huang and Xiangfeng Duan, Nature Reviews Materials, 1, 16042(2016).

\bibitem{QianNiu08} Wang Yao, Di Xiao, and Qian Niu, Phys. Rev. B 77, 235406(2008).

\bibitem{TonyFHeinz12} Kin Fai Mak, Keliang He, Jie Shan and Tony F. Heinz, Nature Nanotechnology, 7, 494(2012).

\bibitem{XiaodongCui12} Hualing Zeng, Junfeng Dai, Wang Yao, Di Xiao and Xiaodong Cui, Nature Nanotechnology, 7, 490(2012).

\bibitem{JiFeng12} Ting Cao, Gang Wang, Wenpeng Han, Huiqi Ye, Chuanrui Zhu, Junren Shi, Qian Niu, Pingheng Tan, Enge Wang, Baoli Liu and Ji Feng, Nature Communications, 3, 887(2012).

\bibitem{XiaodongXu13} Aaron M. Jones, Hongyi Yu, Nirmal J. Ghimire, Sanfeng Wu, Grant Aivazian, Jason S. Ross, Bo Zhao, Jiaqiang Yan, David G. Mandrus, Di Xiao, Wang Yao and Xiaodong Xu, Nature Nanotechnology, 8, 634(2013).

\bibitem{WangYao14} Hongyi Yu, Gui-Bin Liu, Pu Gong, Xiaodong Xu and Wang Yao, Nature Communications, 5, 3876(2014).

\bibitem{TingYu15} Mustafa Eginligil, Bingchen Cao, Zilong Wang, Xiaonan Shen, Chunxiao Cong, Jingzhi Shang, Cesare Soci and Ting Yu, Nature Communications, 6, 7636(2015).

\bibitem{WenYuShan15} Wen-Yu Shan, Jianhui Zhou, and Di Xiao, Phys. Rev. B 91, 035402(2015).

\bibitem{Bertoni16} R. Bertoni, C.W. Nicholson, L. Waldecker, H. H$\ddot{u}$bener, C. Monney, U. De Giovannini, M. Puppin, M. Hoesch, E. Springate, R.T. Chapman, C. Cacho, M. Wolf, A. Rubio, and R. Ernstorfer, Phys. Rev. Lett. 117, 277201(2016).

\bibitem{Rodrigo15} Rodrigo A. Muniz and J. E. Sipe, Phys. Rev. B 91, 085404(2015).

\bibitem{Arzate16} N. Arzate, Bernardo S. Mendoza, R. A. V$\acute{a}$zquez-Nava, Z. Ibarra-Borja, and M. I. $\acute{A}$lvarez-N$\acute{u}\tilde{n}$ez, Phys. Rev. B 93, 115433(2016).

\bibitem{GuiBinLiu13} Gui-Bin Liu, Wen-Yu Shan, Yugui Yao, Wang Yao, and Di Xiao, Phys. Rev. B 88, 085433(2013).

\bibitem{ShiangFang15} Shiang Fang, Rodrick Kuate Defo, Sharmila N. Shirodkar, Simon Lieu, Georgios A. Tritsaris, and Efthimios Kaxiras, Phys. Rev. B 92, 205108(2015).

\bibitem{NataliaCort19} Natalia Cort$\acute{e}$s, O. $\acute{A}$valos-Ovando, L. Rosales, P. A. Orellana, and S. E. Ulloa, Phys. Rev. Lett. 122, 086401(2019).



\bibitem{KaiChang06} J. S. Sheng and Kai Chang, Phys. Rev. B 74, 235315(2006).

\bibitem{KaiChang11} Kai Chang and Wen-Kai Lou, Phys. Rev. Lett. 106, 206802(2011).

\bibitem{WenKaiLou11} Wen-Kai Lou, Fang Cheng, and Jun Li, J. Appl. Phys., 110, 093714(2011).

\bibitem{KaiChang14} Chunxu Bai, Yonglian Zou, Wen-Kai Lou, and Kai Chang, Phys. Rev. B 90, 195445(2014).



\bibitem{luo17} Ma Luo and Zhibing Li, Phys. Rev. B 96, 165424(2017).



\bibitem{Jungwirth16} T. Jungwirth, X. Marti, P. Wadley and J. Wunderlich, Nature Nanotechnology, 11, 231每241(2016).

\bibitem{Baltz18} V. Baltz, A. Manchon, M. Tsoi, T. Moriyama, T. Ono, and Y. Tserkovnyak, Rev. Mod. Phys. 90, 015005(2018).

\bibitem{Zelezny18} J. Zelezny, P. Wadley, K. Olejnik, A. Hoffmann and H. Ohno, Nature Physics, 14, 220每228(2018).

\bibitem{BGPark11} B. G. Park, J. Wunderlich, X. Marti, V. Holy, Y. Kurosaki, M. Yamada, H. Yamamoto, A. Nishide, J. Hayakawa, H. Takahashi, A. B. Shick and T. Jungwirth, Nature Materials, 10, 347每351(2011).

\bibitem{LeiXu18} Lei Xu, Ming Yang, Lei Shen, Jun Zhou, Tao Zhu, and Yuan Ping Feng, Phys. Rev. B 97, 041405(R)(2018).

\bibitem{CastroNeto09} A. H. Castro Neto, F. Guinea, N. M. R. Peres, K. S. Novoselov, and A. K. Geim, Rev. Mod. Phys. 81, 109(2009).








\bibitem{Inglot14} M. Inglot, V. K. Dugaev, E. Ya. Sherman, and J. Barnas, Phys. Rev. B 89, 155411 (2014).

\bibitem{Rioux14} J. Rioux and G. Burkard, Phys. Rev. B 90, 035210 (2014).

\bibitem{Inglot15} M. Inglot, V. K. Dugaev, E. Ya. Sherman, and J. Barnas, Phys. Rev. B 91, 195428 (2015).

\bibitem{Kaladzhyan15} V. Kaladzhyan, P. P. Aseev, and S. N. Artemenko, Phys. Rev. B 92, 155424 (2015).

\bibitem{Reinaldo17} Reinaldo Zapata-Pena, Bernardo S. Mendoza, and Anatoli I. Shkrebtii, Phys. Rev. B 96, 195415(2017).




\bibitem{CLKane051} C. L. Kane and E. J. Mele, Phys. Rev. Lett. 95, 226801(2005).
%

\bibitem{CLKane052} C. L. Kane and E. J. Mele, Phys. Rev. Lett. 95, 146802(2005).
%

\bibitem{TsungWeiChen11} Tsung-Wei Chen, Zhi-Ren Xiao, Dah-Wei Chiou, and Guang-Yu Guo, Phys. Rev. B 84, 165453(2011).





\bibitem{ChangYuHou07} Chang-Yu Hou, Claudio Chamon, and Christopher Mudry, Phys. Rev. Lett. 98, 186809(2007).

\bibitem{ClaudioChamon08} Claudio Chamon, Chang-Yu Hou, Roman Jackiw, Christopher Mudry, So-Young Pi, and Gordon Semenoff, Phys. Rev. B 77, 235431(2008).

\bibitem{DoronLBergman13} Doron L. Bergman, Phys. Rev. B 87, 035422(2013).

\bibitem{GianlucaGiovannetti15} Gianluca Giovannetti, Massimo Capone, Jeroen van den Brink, and Carmine Ortix, Phys. Rev. B 91, 121417(R)(2015).

\bibitem{EliasAndrade19} Elias Andrade, Ramon Carrillo-Bastos, and Gerardo G. Naumis, Phys. Rev. B 99, 035411(2019).

\bibitem{YangLi18} Yang Li, Xue-Yin Sun, Cheng-Yan Xu, Jian Cao, Zhao-Yuan Sunb and Liang Zhen, Nanoscale, 10, 23080(2018).


\bibitem{Baettig05} Pio Baettig, Claude Ederer, and Nicola A. Spaldin, Phys. Rev. B 72, 214105(2005).

\bibitem{Neaton06} J. B. Neaton, C. Ederer, U. V. Waghmare, N. A. Spaldin, and K. M. Rabe, Phys. Rev. B 71, 014113(2005).

\bibitem{Albrecht10} D. Albrecht, S. Lisenkov, Wei Ren, D. Rahmedov, Igor A. Kornev, and L. Bellaiche, Phys. Rev. B 81, 140401(R)(2010).

\bibitem{ZhenhuaQiao14} Zhenhua Qiao, Wei Ren, Hua Chen, L. Bellaiche, Zhenyu Zhang, A. H. MacDonald, and Qian Niu, Phys. Rev. Lett. 112, 116404(2014).

\bibitem{ZYZhu11} Z. Y. Zhu, Y. C. Cheng, and U. Schwingenschl$\ddot{o}$gl, Phys. Rev. B 84, 153402(2011).

\bibitem{DiXiao12} Di Xiao, Gui-Bin Liu, Wanxiang Feng, Xiaodong Xu, and Wang Yao, Phys. Rev. Lett. 108, 196802(2012).



\bibitem{vasp001} P. E. Bl$\ddot{0}$chl, Phys. Rev. B 50, 17953(1994).

\bibitem{vasp002} G. Kresse and J. Furthmuller, Phys. Rev. B 54, 11169(1996).

\bibitem{vasp003} J. P. Perdew, J. A. Chevary, S. H. Vosko, K. A. Jackson, M. R. Pederson, D. J. Singh, and C. Fiolhais, Phys. Rev. B, 46, 6671 (1992).


\bibitem{RiccardoPisoni19} Riccardo Pisoni, Tim Davatz, Kenji Watanabe, Takashi Taniguchi, Thomas Ihn, and Klaus Ensslin, Phys. Rev. Lett. 123, 117702(2019).



\bibitem{CCardoso18} C. Cardoso, D. Soriano, N. A. Garcia-Martinez, and J. Fernandez-Rossier, Phys. Rev. Lett. 121, 067701(2018).

\bibitem{maluo19} Ma Luo, Phys. Rev. B 99, 165407(2019).

\bibitem{maluo16} Ma Luo and Zhibing Li, Phys. Rev. B 94, 235435(2016).

\bibitem{XiaofeiLiu13} Xiaofei Liu, Tao Xu, Xing Wu, Zhuhua Zhang, Jin Yu, Hao Qiu, Jin-Hua Hong, Chuan-Hong Jin, Ji-Xue Li, Xin-Ran Wang, Li-Tao Sun and Wanlin Guo, Nature Communications, 4, 1776(2013).




\end{thebibliography}
\end{document}